 \documentclass[preprint]{aastex}

\newbox\grsign \setbox\grsign=\hbox{$>$} \newdimen\grdimen
\grdimen=\ht\grsign
\newbox\simlessbox \newbox\simgreatbox \newbox\simpropbox
\setbox\simgreatbox=\hbox{\raise.5ex\hbox{$>$}\llap
     {\lower.5ex\hbox{$\sim$}}}\ht1=\grdimen\dp1=0pt
\setbox\simlessbox=\hbox{\raise.5ex\hbox{$<$}\llap
     {\lower.5ex\hbox{$\sim$}}}\ht2=\grdimen\dp2=0pt
\setbox\simpropbox=\hbox{\raise.5ex\hbox{$\propto$}\llap
     {\lower.5ex\hbox{$\sim$}}}\ht2=\grdimen\dp2=0pt
\def\simgreat{\mathrel{\copy\simgreatbox}}
\def\simless{\mathrel{\copy\simlessbox}}

\shorttitle{Variable X-ray Absorption in the Seyfert 2 Galaxy Mrk 348}
\shortauthors{Smith et al.}

\begin{document}

\title{Variable X-ray Absorption in the Seyfert 2 Galaxy Mrk 348}

%% Use \author, \affil, and the \and command to format
%% author and affiliation information.
%% Note that \email has replaced the old \authoremail command
%% from AASTeX v4.0. You can use \email to mark an email address
%% anywhere in the paper, not just in the front matter.
%% As in the title, you can use \\ to force line breaks.

\author{David A. Smith\altaffilmark{1}}
\affil{Laboratory for High Energy Astrophysics, NASA/GSFC, Code 662,
	Greenbelt, MD 20771}

\author{Ioannis Georgantopoulos}
\affil{National Observatory of Athens, Lofos Koufou, Palaia Penteli,
	15236 Athens, Greece}

\and 

\author{Robert S. Warwick}
\affil{Department of Physics and Astronomy, University of Leicester,
	Leicester LEI 7RH, UK}

%% Notice that each of these authors has alternate affiliations, which
%% are identified by the \altaffilmark after each name.  Specify alternate
%% affiliation information with \altaffiltext, with one command per each
%% affiliation.

\altaffiltext{1}{also Astronomy Department, University of Maryland, College 
	Park, MD 20742}

\begin{abstract}

We present RXTE monitoring observations of the Seyfert 2 galaxy Mrk
348 spanning a 6 month period. The time-averaged spectrum in the 3--20
keV band shows many features characteristic of a Compton-thin Seyfert
2 galaxy, namely a hard underlying power-law continuum ($\Gamma
\approx 1.8$) with heavy soft X-ray absorption ($N_{H} \sim
10^{23}$~cm$^{-2}$) plus measureable iron K$\alpha$ emission
(equivalent width $\sim 100$ eV) and, at high energy, evidence for a
reflection component ($R \simless 1$). During the first half of the
monitoring period the X-ray continuum flux from Mrk 348 remained
relatively steady. However this was followed by a significant
brightening of the source (by roughly a factor of 4) with the fastest
change corresponding to a doubling of its X-ray flux on a timescale of
about 20 days. The flux increase was accompanied by a marked softening
of X-ray spectrum most likely attributable to a factor $\sim 3$
decline in the intrinsic line-of-sight column density.  In contrast
the iron K$\alpha$ line and the reflection components showed no
evidence of variability.  These observations suggest a scenario in
which the central X-ray source is surrounded by a patchy distribution
of absorbing material located within about a light-week of the nucleus
of Mrk 348.  The random movement of individual clouds within the
absorbing screen, across our line of sight, produces substantial
temporal variations in the measured column density on timescales of
weeks to months and gives rise to the observed X-ray spectral
variability. However, as viewed from the nucleus the global coverage
and typical thickness of the cloud layer remains relatively constant.
 
\end{abstract}


\section{Introduction} \label{sec-intro}

The discovery of ``hidden'' broad-line regions (BLRs) in several
Seyfert 2 galaxies \citep{am85,mg90,tra92} has generated much interest
in unified theories, in which the viewing orientation explains many of
the observed differences between type 1 and 2 Seyfert galaxies
(Antonucci 1993 and references therein). Such theories posit that, in
both types of Seyfert galaxy, a luminous central source (presumably an
accreting supermassive blackhole) is surrounded by a equatorial
distribution of matter in the form of either a swollen accretion disk
or a molecular torus \citep{kb86}. A type 1 nucleus is then observed
only if the viewing direction is sufficiently close to the axis of the
system so as to fall within the opening angle of the putative
torus. More inclined lines of sight necessarily intercept the
disk/torus material in which case we see an ``obscured'' type 2
system.  However, a caveat is that a small fraction of the nuclear
flux can be scattered into our line of sight by highly ionized
material filling the ``hole'' of the torus, thus explaining the
detection of broad optical lines in polarized light and also weak UV
and soft X-ray emission even in Seyfert 2s where our direct view of
the nucleus is completely blocked.

X-ray measurements of intrinsic column densities of $\simgreat
10^{22}$ cm$^{-2}$ in many Seyfert 2 galaxies
\citep{awa91,sd96,tur97a} lend support to the above unification
picture. For column densities $\simless 10^{24}$ cm$^{-2}$, X-rays can
penetrate the torus leading to the detection of an absorbed power-law
continuum together with a iron K$\alpha$ fluorescence line at 6.4~keV.
However for column densities $\simgreat 10^{24}$ cm$^{-2}$, even hard
X-radiation is seen only indirectly via scattering (as in
Compton-thick Seyfert 2s such as NGC 1068).  X-rays may be electron
scattered by highly ionized matter extending along the axis of the
system or Compton reflected by almost neutral material located at the
inner edge of the torus.  In either case, the scattering material is
exposed to the intense nuclear flux and may produce significant line
emission and, in circumstances where the direct continuum is
completely blocked leaving only a baseline of the much weaker
scattered continuum, such lines can have very high equivalent widths
\citep{ghi94,kro94}.
  
The spectral indices found in Seyfert galaxies are distributed around
a single value of $\Gamma \approx 1.9$ \citep{np94,sd96}.  The best
explanation for this is that mildly relativistic thermal electrons or
highly relativistic non-thermal electrons Compton scatter UV photons
into the X-ray energy range (e.g. Svensson 1996 and references
therein).  The rollover seen in X-ray data above 100 keV has focussed
most attention on thermal models \citep{zdz95,gon96}.  A popular
scenario is the two-phase disk-corona model, in which a hot X-ray
corona is located above a cold UV emitting accretion disk
\citep{hm91,hm93}.  The UV seed photons for Compton cooling of the
energetic electrons are produced from reprocessing of the hard X-ray
spectrum in the accretion disk.  With complete feedback, the
approximate equipartition between the soft disk and hard X-ray
luminosities leads naturally to $\Gamma \simgreat 1.9$ for a wide
range of optical depths (see also Stern et al. 1995).  However, for
sources with spectra flatter than $\Gamma \approx 1.9$, the
disk-corona model of \citet{hm91,hm93} must be modified so that some
of the soft seed photons escape without being intercepted by the
corona, thus making for harder spectra as fewer photons are available
for Compton scattering \citep{haa94}.

X-ray variability studies of Seyfert 2 nuclei can provide important
clues to the processes by which X-rays are produced as well as to the
geometry of the circumnuclear matter.  However, variability has been
reported in only a few objects.  One example is Mrk 3, where the hard
X-ray emission decreased by a factor of two during a period of 3.6
years \citep{awa90,mar92,iwa94}.  Recent observations with \emph{RXTE}
have shown variability to exist on time-scales of weeks \citep{geo99}.
The shape of the spectrum suggests that \emph{all} of the variability
can be attributed to changes in the direct nuclear flux \citep{cap99}.
Short term variability (on time-scales of hours) has been detected in
only one object, NGC 4945 \citep{iwa93,gua00}.  These observations
suggest that a parsec scale, geometrically thick molecular torus
cannot be responsible for the bulk of the X-ray absorption
\citep{mad00}.

Spectral variability has also been claimed in a couple of Seyfert 2
galaxies.  For example, the absorption column in NGC 7582 increased by
$N_{\rm H} \sim 4 \times 10^{22}$ cm$^{-2}$ between \emph{ASCA}
observations taken two years apart \citep{xue98}.  A similar variation
in the absorber was suggested by \citet{war93}, based on a comparison
between \emph{Einstein}, \emph{EXOSAT}, and \emph{Ginga} observations.
More recent observations with \emph{BeppoSAX} of this particular
galaxy have shown that the absorber may have a complex spatial
structure with a large column density ($N_{\rm H} \sim 10^{24}$
cm$^{-2}$) covering 60\% of hard X-ray continuum, and a smaller, but
variable column completely covering the source \citep{tur00}.  As a
further example, in NGC 7172, the spectral index of the power law
continuum decreased by $\Delta \Gamma \approx 0.3$ between
\emph{Ginga} and \emph{ASCA} observations taken almost 4 years apart
\citep{ryd97}.  These authors attribute the decrease to an intrinsic
change in the spectral index although, given the narrow bandpass of
\emph{ASCA}, it is possible that the spectrum is affected by a
complex absorber and/or Compton reflection.

The launch of the \emph{RXTE} satellite is conducive to the systematic
study of variability in Seyfert 2 galaxies, because of the large
effective area and broad band coverage afforded by the detectors, and
the flexible scheduling allowed by the mission.  We have conducted a
six month monitoring campaign with \emph{RXTE} of a few Seyfert 2
galaxies previously observed by the \emph{Ginga} and \emph{ASCA}
satellites.  The results of our observations of Mrk 3 have been
reported elsewhere \citep{geo99}.  Here we present our results on Mrk
348.

Mrk~348 is a nearby ($z=0.015$; de Vaucouleurs et al. 1991) Seyfert 2
galaxy for which there is evidence of a broad (FWHM $\sim 7400$
km~s$^{-1}$) H$\alpha$ line component in polarized light \citep{mg90}.
In terms of current unification schemes for Seyfert galaxies (see
above) this can be taken as clear evidence for the presence of an
obscured Seyfert 1 nucleus, confirmation of which was provided by the
\emph{Ginga} detection of an absorbed ($N_{\rm H} =
10^{23.10\pm0.08}$ cm$^{-2}$) hard X-ray source (with $\Gamma = 1.68
\pm 0.17$) coincident with this galaxy \citep{war89}.  This heavy
obscuration in the nucleus of Mrk 348 may also be responsible for the
bi-canonical nature of its ionizing continuum \citep{mul96,sim96}.

There was marginal evidence in the \emph{Ginga} spectrum for an iron
K$\alpha$ emission line at 6.4 keV of equivalent width $130\pm130$ eV.
An iron K$\alpha$ emission line of equivalent width $113\pm56$ eV was
subsequently detected with \emph{ASCA} \citep{net98}.  In addition to
iron K$\alpha$, \emph{ASCA} detected a prominent S {\sc xiv}--{\sc
xv\/} line of large equivalent width ($600\pm300$ eV).  A weak power
law continuum was detected at soft X-ray energies in observations made
with the \emph{ROSAT} Position Sensitivity Proportional Counter (PSPC)
($\Gamma = 2.40^{+0.58}_{-0.42}$; Mulchaey et al. 1993) and
\emph{ASCA} ($\Gamma = 2.77\pm0.87$; Netzer et al. 1998).  It is most
likely that this weak soft X-ray emission is dominated by
electron-scattered nuclear flux, with perhaps some contribution from
spatially extended thermal emission from a hot wind (e.g. Krolik \&
Vrtilek 1984).  The 2--10 keV luminosity measured in the \emph{Ginga}
observation was $1.2 \times 10^{43}$ erg s$^{-1}$ ($H_{\circ} = 50;
q_{\circ} = 0.5$), which is roughly a factor 3 higher than that
subsequently recorded by \emph{ASCA}.

\section{Observations and Data Analysis} \label{sec-obs}

The \emph{Rossi X-ray Timing Explorer (RXTE)} satellite made twelve
separate observations of Mrk~348 during the period December 29, 1996
to July 12, 1997, with the intervals between the observations arranged
so that the X-ray variability would be sampled on timescales of days,
weeks, and months (Table~\ref{tbl-1}). The observation durations were
typically in the range 2500--5000 seconds. The instruments on board
\emph{RXTE} are the Proportional Counter Array (PCA; Glasser, Odell
\& Seufert 1994), the High Energy X-ray Timing Experiment (HEXTE;
Gruber et al. 1996) and the All-Sky Monitor (ASM; Levine et
al. 1996). However, in this paper we consider only the PCA data, since
Mrk~348 was not detected by the other instruments.  The PCA consists
of five sealed proportional counter units (PCUs), with a total
collecting area of $\sim 6500$~cm$^{2}$ \citep{jah96}. The main
detector volume of each PCU contains a 90\% xenon plus 10\% methane
gas mixture, and three main layers of anode wires, connected in such a
way that two chains of events are produced per layer
\citep{gla94}. The detectors are sensitive to X-rays in the 2--60~keV
range, and have an energy resolution equal to $\Delta E/E \sim 18\%$
(FWHM) at 6~keV. The field of view of each PCU is restricted to
$1^{\circ}$ (FWHM) by a hexagonal, beryllium copper collimator
\citep{jah96}.

Data reduction was carried out using the latest version of the {\sc
ftools\/} software (version 4.2) together with the standard-2 data
files. The standard-2 mode data are the most useful when analysing
relatively faint sources, since spectra with full layer identification
are accumulated every 16~seconds from ``good'' xenon events
(i.e. those which survive background rejection). For every standard-2
mode data file, a background file is generated using {\sc
pcabackest\/} and the ``L7\_240'' faint source model files. The
internal background is correlated with the ``L7'' rate, which is the
sum of the seven two-fold coincidence rates present in the standard-2
mode data files. Most of the internal background can be estimated by
matching conditions in the standard-2 files to those in the ``L7''
model file. However, there is a residual background component
associated with activation induced in the spacecraft as it passes
through the South Atlantic Anomaly (SAA). In estimating the background
due to activation, which has a half-life of 240 minutes, the HEXTE
particle monitor rate is integrated, with an exponential decay term,
over the most recent SAA passages. This integrated rate is then
multiplied with the activation spectrum in the ``240'' model file to
give the total activation component.

Light curves and spectra were extracted from both the standard-2 mode
and background data file, using the standard data selection criteria
for faint sources. Only front layer light curves and spectra are
considered, as these data have the largest signal-to-noise ratio. The
background dominates above $20~\rm keV$, so data above this energy are
not considered further, and the low energy cut-off for PCA data is
2~keV. PCU3 and PCU4 are occasionally switched off because residue
collects on the anodes, so increasing the electric field causing
discharge, and therefore the analysis is restricted to data from the
first 3~PCUs. Deadtime corrections were not applied, because the
correction is very small ($\sim 1\%$) even when the count rate per PCU
reaches $1000 \rm~count~s^{-1}$. The individual response matrices for
PCU0, PCU1, and PCU2 are co-added prior to spectral fitting.

\section{Timing Analysis} \label{sec-timing}

The background-subtracted count rates measured in the 2--6~keV band
(where the flux is significantly affected by absorption), 6--10~keV
band (where the underlying power-law probably dominates the flux), and
10--20~keV band (where Compton reflection may become important) are
shown in Figure~\ref{fig1}.  In all three energy ranges, variability
becomes apparent in the second half of the monitoring period and is
most evident towards the very end of the campaign. Specifically, the
count rate doubles over a period as short as 20~days.  The rms
variability, after correcting for uncertainties in the background
model\footnote{\tt http://lheawww.gsfc.nasa.gov/\~\,$\!$dasmith/systematics\_000222/systematics\_000222.html}, amounts to 61, 48, and 42\% for the 2--6, 6--10 and 10--20~keV bands respectively.

The strong correlation between the signal in the three bands suggests
that a single, broad-band spectral component may be responsible for
much of the observed variability. The spectral changes implied by the
decline in the variability amplitude with energy might result from
spectral variations in this component or, alternatively, indicate the
presence of two or more spectral components with different temporal
behaviour.  Further illustration of the X-ray spectral variability is
provided by the light curves of the 10--20~keV/6--10~keV and
6--10~keV/2--6~keV hardness ratios, denoted HR1 and HR2 respectively,
both of which show a spectral softening as the source intensity
increases (see Figure~\ref{fig2}).

\section{Spectral Fitting} \label{sec-spectral}

To improve the signal-to-noise ratio, individual spectra with similar
hardness ratio values were combined (i.e., observations 1 \& 2; 3 \&
4; 5 \& 6; 7, 8 \& 9; 10 \& 11; 12) to give a total of six spectral
datasets as input to the spectral fitting analysis, which is based on
the {\sc xspec v10.0\/} software package \citep{arn96}.

Above 25 keV, where no signal is expected from Mrk 348 in the front
layer, the observed count rates are within 1\% of the predicted
background count rate. However, below 6 keV, where the mid- and
bottom-layer counts rates should be close to the background level, a
small deficit is observed in three spectra, which can be removed by
decreasing the background by $\leq 5$\%. Hence, the systematic error
in the background is considered to be $\leq 5$\% in three of the six
datasets, and $\leq 1$\% elsewhere.  Following \citet{lei99}, an
estimate of the uncertainties due to the systematic error in the
background can be obtained by altering the normalization of the
background spectrum by $\pm 1$\% or $\pm 5$\%.  These are shown as
additional errors on the spectral parameters.

Fluctuations in the Cosmic X-ray Background (CXB) are also a
significant source of uncertainty in the PCA spectra of faint
sources. Based on results obtained with the \emph{Ginga} Large Area
Counter instrument, the $1\sigma$ fluctuation spectrum is best
described by a power-law of energy index $\Gamma = 1.8$ and
normalization $\pm 2.0 \times 10^{-4}$ photons cm$^{-2}$ sec$^{-1}$
keV$^{-1}$ \citep{but97,rey99}. Hence in our analysis of the X-ray
spectrum of Mrk 348, a fluctuation spectrum was included as a separate
model component, thus allowing the data themselves to constrain the
level of the fluctuation required. This technique was, for example,
previously used by \citet{sd96} in their analysis of \emph{Ginga}
Seyfert~2 spectra.

As a starting point in the spectral analysis all six spectra were
fitted by model consisting of a power-law continuum (photon spectral
index $\Gamma$) absorbed by an intrinsic column density ($N_{\rm H}$),
where the atomic cross-sections and abundances for the latter were
taken from \citet{bm92} and \citet{ag89}, respectively. We also
included a narrow ($\sigma_{\rm K\alpha} = 0.1$~keV) Gaussian emission
line to represent the iron K$\alpha$ emission.  Initially, only the
power-law normalization was allowed to vary across the six spectra
with the line energy fixed at $6.4$~keV, the value for neutral
iron. This model (model~1) provides an unacceptable fit with $\chi^{2}
= 484.2$ for $= 302$ degrees of freedom (d.o.f.), the best-fitting
parameters being $\Gamma = 1.54^{+0.03}_{-0.04}$ and $N_{\rm H} =
10.1^{+0.5}_{-0.6} \times 10^{22}$~cm$^{-2}$ (the errors, here and
elsewhere in this paper, are 90\% confidence for one interesting
parameter, $\Delta \chi^{2} = 2.7$).  Given the spectral variability
evident in the data (see section~\ref{sec-timing}), it is not
surprising that this model rather poorly describes the observed
spectra.

The next step was to allow either the column density (model 2) or the
photon index (model 3) to vary across the six spectra. The variable
column density description provides an excellent fit, with $\Gamma =
1.58\pm0.03$ and $N_{\rm H}$ in the range ($8.6^{+0.8}_{-0.6}$ to
$31.8^{+5.5}_{-4.6}$) $\times 10^{22}$ cm$^{-2}$ ($\chi^{2} = 202.1$,
297 d.o.f.).  Consistent with the observed hardness ratio variations,
the column density falls as the power-law normalization (and hence the
intrinsic flux) increases.  The variable index model provides a worse
fit, but nevertheless one which is still excellent, with $\Gamma$ in
the range $0.92^{+0.09}_{-0.10}$ to $1.68^{+0.05}_{-0.04}$ and $N_{\rm
H} = 9.8^{+0.5}_{-0.5} \times 10^{22}$~cm$^{-2}$ ($\chi^{2} = 221.5$,
297 d.o.f.).  In this case there is a strong correlation between
photon index and intrinsic flux, with a steeper spectral index being
preferred in the brighter source states (see Table~\ref{tbl-2} for
details of both model 2 and model 3 spectral fitting results).  The
2--10 keV luminosity measured in the \emph{RXTE} observations was
($0.8$--$3.4$) $\times 10^{43}$ ergs s$^{-1}$ ($H_{\circ} = 50;
q_{\circ} = 0.5$), which implies a peak intensity a factor 3 higher
than that recorded by \emph{Ginga}.

In the variable column density description (model 2), the {\it
unabsorbed\/} iron K$\alpha$ emission line flux was $2.1^{+0.8}_{-0.7}
\times 10^{-5}$ photon s$^{-1}$ cm$^{-2}$, yielding an equivalent
width, with respect to the {\it absorbed\/} continuum, in the range
40--170 eV.  This is consistent with the emission line fluxes
determined from earlier \emph{Ginga} ($5.0\pm2.2 \times 10^{-5}$
photon s$^{-1}$ cm$^{-2}$; Smith \& Done 1996) and \emph{ASCA}
($1.4\pm0.7 \times 10^{-5}$ photon s$^{-1}$ cm$^{-2}$; Netzer, Turner
\& George 1998) observations, although the uncertainties in the
individual measurements are rather large.  Assuming a spherical
distribution of gas, the equivalent width expected from the measured
column density would be in the range 40--130 eV \citep{lc93},
comparable to that observed.  We note that the best-fit line energy is
$6.15\pm0.14$ keV, which is below that expected for cold iron.
However, we suspect this discrepancy is due to uncertainties in the
instrument response around the Xe L edge at $\sim 5$ keV.  Finally, if
the line normalization is allowed to vary between the six spectra, the
reduction in $\chi^{2}$ is $\sim 3$, which is not statistically
significant (on the basis of the F-test for five additional free
parameters; Bevington \& Robinson 1992).  Although this lack of line
variability is intriguing, the rather large uncertainties in the
individual measurements and coarse sampling do not exclude the
possibility that the line flux tracks the continuum on timescales as
short as hours, as would be expected if most of the line was produced
in a putative accretion disk.

While the spectra fits confirm the presence of spectral variability,
the origin of the changes in either column density and/or photon index
variations is not distinguishable. Therefore, as an additional test,
the data were separated into two spectra on the basis of the flux
change that occurred during the latter part of the monitoring
campaign. Thus, data from the first six observations were combined
into a ``low state'' spectrum, while data from the last six spectra
comprise a ``high state'' spectrum. These two spectra were then fitted
to a spectral model in which {\it both\/} the column density and
photon index were allowed to vary. The projected confidence regions in
the $\Gamma$--$N_{\rm H}$ plane for this model are shown in
Figure~\ref{fig3}. Although photon index variations cannot be
excluded, a significant change in the column density is required by
the data. Hereafter we take the view that, on the basis of these very
simple spectral models, the spectral variability is best assigned
largely to variations in the column density.

Although the above absorbed power-law model provides an excellent fit 
to the combined \emph{RXTE} datasets, we next consider what constraints, 
if any, may be placed on slightly more sophisticated spectral descriptions,
such as models including either a Compton reflection component or a 
partially covered source. Specifically we considered the possibility that 
some of the observed spectral variability may arise due to the presence
of an extra component with a delayed temporal response relative to 
the direct continuum. 

If the disk-corona model described in \S 1 is correct, then a
significant fraction of the hard X-ray flux should be reprocessed in
the accretion disk.  The anticipated spectral signatures from such
reprocessing includes an iron K$\alpha$ fluorescence line and Compton
reflection of the hard continuum \citep{gf91,mat91}.  These features
have previously been identified in many Seyfert galaxies, and may also
be present in the spectrum of Mrk 348.  Motivated by this prediction,
the data were fitted to a model consisting of an absorbed power-law
continuum with {\it absorbed} Compton reflection from neutral, solar
abundant material, as implemented in the {\sc pexrav\/} model
\citep{mz95} plus a narrow Gaussian line representing iron K$\alpha$
emission (model 4).  The inclination (to our line of sight) of the
reflecting material was fixed at the default value ($i = 60^{\circ}$),
since the reflection spectrum below 20 keV is relatively independent
of the inclination angle. The shape of the incident continuum above 20
keV is important, even when considering measurements below this energy
because of the effects of Compton down-scattering; the power-law
continuum was therefore exponentially cut-off with an e-folding energy
of 150 keV, similar to that observed in other Seyfert galaxies
\citep{zdz95,gon96}.

The reflected spectrum should track the continuum on timescales of
weeks or less, if it originates in the putative disk.  Therefore, the
relative normalization between the direct and reflected continuum was
initially tied to a single value for all six spectra (a value of $\cal
R$~$= 1$ is expected from a flat $\Omega/2\pi = 1$ geometry and
isotropic emission).  In this case the best-fitting parameter values
are $\Gamma = 1.69^{+0.07}_{-0.07}$, $\cal R$~$= 0.5^{+0.2}_{-0.1}$
and $N_{\rm H}$ in the range ($9.1 \pm 0.6$ to $31.9^{+5.4}_{-4.3}$)
$\times 10^{22}$ cm$^{-2}$ ($\chi^{2} = 193.4$; $d.o.f. = 296$; see
Table~\ref{tbl-3}).  This model provides a significantly ($> 99$\%)
better fit to the data (on the basis of a F-test for one additional
free parameter), than our earlier preferred model (model 2).  The iron
K$\alpha$ emission line flux was $1.6^{+0.8}_{-0.7} \times 10^{-5}$
photons s$^{-1}$ cm$^{-2}$, which yields an equivalent width in the
range 40--120 eV, compared with the $\sim 50$ eV equivalent width
expected from the reflection continuum \citep{gf91,mat91,ghi94}.
Thus, for the brighter source states, reflection could produce all of
the observed line emission.  When the relative strength of the
reflection continuum is allowed to vary between the observations,
there is a strong correlation between its strength and the intensity
of the source, with a stronger reflection continuum being preferred
for the weaker source states.  However, the reduction in $\chi^{2}$ of
$\Delta \chi^{2} \approx 5$ is insignificant (on the basis of a F-test
for six additional free parameters) and, therefore, we cannot draw any
firm conclusions.

If the molecular torus is optically thick, then Compton reflection
originating at the (far-side) inner walls of the torus may be seen
directly, rather than through the gas column responsible for the low
energy absorption (depending, of course, on the torus geometry and
view angle). Such reflection would not be expected to vary on
timescales shorter than about a year, and could in principle lead to a
net softening of the spectrum as the source intensity increases.  Such
reflection could also contribute significantly to a non-varying
component of the iron K$\alpha$ emission line.  To test this
possibility, the data were fitted to a model including an absorbed
power-law continuum, an {\it unabsorbed and non-variable} Compton
reflection component, and a narrow Gaussian emission line (model 5).
The Compton reflection spectrum was the same as that used for model 4.
The best-fit values are $\Gamma = 1.65 \pm 0.05$ and $N_{\rm H}$ in
the range ($9.8^{+0.9}_{-0.7}$ to $32.4^{+6.8}_{-6.2}$) $\times
10^{22}$ cm$^{-2}$ ($\chi^{2} = 188.5$; $d.o.f = 296$; see
Table~\ref{tbl-3}).  The improvement in $\chi^{2}$ over that for the
variable absorber model is significant at $> 99$\% (on the basis of a
F-test for one additional term).  The relative normalization between
the direct and reflected continua varies from $\cal R$~$\approx 0.3$
to $\cal R$~$\approx 0.8$ as the level of the former changes.  The
iron K$\alpha$ emission line flux was $1.6 \pm 0.7 \times 10^{-5}$
photons s$^{-1}$ cm$^{-2}$, and yields an equivalent width in the
range 30--120 eV, comparable to the 30--100 eV range expected from
reflection continuum alone.

Recently, evidence has emerged for Compton-thick material which
partially, or fully covers the central nucleus in several Seyfert 2
galaxies (NGC 4945, Iwasawa et al. 1993, Done, Madejski \& Smith 1996;
IRAS 04575-7537, Vignali et al. 1998; Mrk 3, Turner et al. 1997b,
Cappi et al. 1999, Georgantopoulos et al. 1999; NGC 7582, Turner et
al. 2000).  In order to test for the presence of Compton-thick
material obscuring the nucleus in Mrk 348, the data were fitted to a
model consisting of a power-law continuum, which is partially covered
by a constant column ($N_{\rm H1}$), and fully covered by a variable
absorber ($N_{\rm H2}$), plus a narrow Gaussian emission line (model
6). This model provides the best fit so far to the data, with $\Gamma
= 1.77^{+0.10}_{-0.05}$, $N_{\rm H1} = (111^{+52}_{-32}) \times
10^{22}$ cm$^{-2}$ covering $0.24^{+0.03}_{-0.07}$ of the source, and
$N_{\rm H2}$ in the range ($9.7^{+0.8}_{-0.6}$ to
$31.9^{+5.4}_{-5.1}$) $\times 10^{22}$ cm$^{-2}$ ($\chi^{2} = 184.9$;
$d.o.f. = 295$; see Table~\ref{tbl-4}).  The iron K$\alpha$ line
equivalent width expected from an obscuring column of $N_{\rm H} \sim
10^{24}$ cm$^{-2}$ would be $\sim 500$~eV \citep{lc93} with respect to
the absorbed continuum, but only $\sim 100$~eV with respect to the
unabsorbed continuum, and so would remain largely undetected in these
data, given the small covering fraction.  We note also, that the
best-fit photon index is $\Gamma \sim 1.8$, which is close to that
found in other Seyfert nuclei \citep{np94,sd96}.

\section{Source Confusion}\label{sec-confusion}

One issue that needs to be addressed is whether confusing sources
within the $1^{\circ}$ field of view of the PCA could have contributed
to the flux and spectral variability noted above. The \emph{ROSAT}
PSPC image of the Mrk 348 field shows many faint soft X-ray sources in
the vicinity of the Seyfert 2 galaxy, but none are bright enough to be
of any concern in the present analysis (except perhaps in the very
unlikely event that one of these sources brightened by an order of
magnitude during the RXTE observations). We note that the LINER 1.9
galaxy, NGC 266 \citep{ho97a}, is situated at a similar redshift
($z=0.016$; Ho et al. 1997b) as Mrk 348 and $23^{\prime}$ from it (at
which point the PCA collimator transmission is roughly 50\% of the
on-axis value; Zhang et al. 1993).  Unfortunately NGC 266 is partially
obscured by a window-support rib in the \emph{ROSAT} PSPC
observation. There is a faint source at the position of NGC 266 in the
equivalent \emph{ASCA} Gas Imaging Spectrometer (GIS) image, but is
$\simgreat 70$ times weaker than Mrk 348 (although there is some
uncertainty in this estimate due to the position of NGC 266 near the
edge of the GIS field of view).  Furthermore, assuming the
H$\alpha$/hard X-ray correlation of \citet{ter00} holds for NGC 266,
its broad H$\alpha$ line emission of $8 \times 10^{40}$ ergs s$^{-1}$
\citep{ho97a} implies a 2--10 keV (unabsorbed) luminosity of order
$10^{41}$ ergs s$^{-1}$.  This again suggests that the contribution of
NGC 266 to the hard X-ray flux measured by \emph{RXTE} is negligible
($\approx 10^{-13}$ ergs s$^{-1}$ cm$^{-2}$ in the 2--10 keV band).
In summary we believe that it is very unlikely that the current X-ray
spectral variability measurements for Mrk 348 are significantly
contaminated by confusing sources in the \emph{RXTE} field of view.

\section{Discussion} \label{sec-discussion}

The shortest timescale on which X-ray continuum variations are
observed in Mrk 348 is $\sim 1$~day, the minimum spacing between
observations.  In terms of the light-travel time this implies a
maximum source size of $r_{\rm s} \simeq 2.5 \times 10^{15}$~cm,
although in reality the X-ray emitting region may be much smaller than
this, (i.e. $\sim 3 \times 10^{14}$~cm if the X-rays are produced
within 10~Schwarzchild radii of a $10^{8}~M_{\odot}$ black hole). In
contrast variations in the intrinsic column density (along the line of
sight to the nucleus of Mrk 348) take place over periods of typically
weeks to months with the largest change, $\Delta N_{\rm H} \sim
10^{23}$~cm$^{-2}$, occurring on a timescale of $\sim 70$ days.

If the observed column changes are interpreted as due to the motion,
across the line of sight, of clouds of gaseous material in orbit
around the central black-hole, then for a transverse cloud velocity of
$v_{\rm cloud} \approx 10,000$~km~s$^{-1}$ (characteristic of extreme
BLR cloud velocities), the predicted timescale for the column change
is $r_{\rm s}/v_{\rm cloud} \simeq 30$~days, which is roughly
consistent with the observations.  Such clouds are located at a
distance of $R = GM/v_{\rm cloud}^{2} \simeq 1.3 \times 10^{16}
M_{8}$~cm, where $M_{8}$ is the black-hole mass in units of
$10^{8}~M_{\odot}$.  We can infer the ionization of the cloud material
via the ionization parameter $\xi$, where $\xi = L_{\rm ion}/n_{\rm e}
R^{2}$.  Here $L_{\rm ion}$ is the ionizing luminosity (measured
between 13.6~eV to 13.6~keV; Kallman \& McCray 1982), $n_{\rm e}$ is
the gas density and $R$ is the distance to the clouds from the
continuum source.  For an ionizing luminosity of $\sim 5 \times
10^{43}$~erg~s$^{-1}$ and $n_{\rm e} \approx 10^{10}$~cm$^{-3}$ (a
typical value for BLR clouds), the ionization parameter is $\xi
\approx 30$. This is consistent with a lightly ionized state for the
iron in the clouds (i.e. below Fe{\sc xv}), but in agreement with the
iron K$\alpha$ emission-line (and K-edge) requirements.  However for a
typical cloud column density of $\Delta N_{\rm H} \sim
10^{23}$~cm$^{-2}$, the implied physical thickness, is only $\Delta
N_{\rm H}/n_{\rm e} \sim 10^{13}$ cm. Since this dimension is much
smaller than the inferred X-ray source size $r_{\rm s}$, the
organisation of clouds into a coherent sheet-like distribution is
required in order to explain the large amplitude $N_{\rm H}$
variations which are observed in Mrk 348 and other sources
(e.g. Yaqoob et al. 1993).  

Alternatively, the increase in the source luminosity may be sufficient
to induce the observed decline in the column density via
photo-ionization.  However, the lack of evidence for highly ionized
iron in Mrk 348 would require a fairly narrow regime for the
ionization parameter.  Such a scenario could be investigated using
photo-ionization codes such as {\sc xstar\/}, but this is beyond the
scope of the present paper.

Within the measurement constraints, the iron K$\alpha$ line flux
remained at a roughly constant level throughout the monitoring
campaign.  This implies that much of the line emission is produced in
material reasonably distant from the nucleus.  A possible location for
this material is in the patchy cloud distribution responsible for the
highly variable intrinsic absorption in this source.  The lack of
significant changes in the line flux then requires both the sky
coverage and typical thickness of the cloud layer, as viewed from the
nucleus, to remain fairly constant (a requirement which need not, of
course, conflict with the fact that temporal variations are observed
along our specific line of sight).  It is also possible that a
contribution to the line emission comes from the illumination of the
(far-side) inner walls of the putative molecular torus (assuming at
least some element of the gaseous medium surrounding the X-ray source
is Thomson thick).  While Compton reflection is not specifically
required by the spectral fitting, the data are consistent with the
reflection expected from a torus subtending a solid angle at the
source of $\Omega/2\pi \approx 0.3$--$0.8$.  If the ionization cones
are collimated by the torus, then the observed half-opening angle of
the cones, of $\theta \sim 45^{\circ}$ \citep{sim96}, corresponds to a
torus solid angle of order $\Omega/2\pi = 2 \cos \theta = 1.4$.  Thus,
in principle the bulk of the line emission could originate in this
way. In this setting, the fact that Mrk 348 is not observed as a
Compton-thick Seyfert 2 is presumably the result of our line of sight
to the nucleus passing across the top of the torus thus avoiding its
most optically thick regions.

Finally, we note that the iron K$\alpha$ emission observed in the
X-ray spectra of many Seyfert~1 galaxies is most naturally associated
with reflection from the inner regions of an accretion disk.  One
signature of such emission is the broadening of the line due to
relativistic effects close to the central massive black-hole
(e.g. Tanaka et al. 1995).  Also, depending on the geometry and
optical depth of the X-ray emitting plasma, variations in the line
flux on timescales of a week or less are predicted. The current
observations provide no evidence for such an origin for the iron
K$\alpha$ emission observed in Mrk 348.  Also an earlier \emph{ASCA}
observation of Mrk 348 was unable to determine whether the iron
K$\alpha$ emission line was significantly broadened \citep{net98}.
Clearly further observations with much higher signal-to-noise and
energy resolution will be needed in order to test for the presence of
an accretion disk in Mrk 348.

\acknowledgments

%% Generally speaking, only the figure captions, and not the figures
%% themselves, are included in electronic manuscript submissions.
%% Use \figcaption to format your figure captions. They should begin on a
%% new page.

\clearpage

%% No more than seven \figcaption commands are allowed per page,
%% so if you have more than seven captions, insert a \clearpage
%% after every seventh one.

%% There must be a \figcaption command for each legend. Key the text of the
%% legend and the optional \label in curly braces. If you wish, you may
%% include the name of the corresponding figure file in square brackets.
%% The label is for identification purposes only. It will not insert the
%% figures themselves into the document.
%% If you want to include your art in the paper, use \plotone.
%% Refer to the on-line documentation for details.

\figcaption[fig1.ps]{Background-subtracted X-ray light curves in the 2-6,
6-10, and 10-20 keV bands. \label{fig1}}

\figcaption[fig2.ps]{Light curves of the 10--20~keV/6--10~keV and 
6--10~keV/2--6~keV hardness ratios, denoted HR1 and HR2 respectively.
\label{fig2}}

\figcaption[fig3.ps]{The projected confidence regions in the
$\Gamma$--$N_{\rm H}$ plane for the high (leftmost contour) and low
state (rightmost contour) data. The contours are $\Delta \chi^{2} =
2.30$, 4.61, and 9.21 (i.e. 68, 90, and 99\% confidence for two
interesting parameters). \label{fig3}}

\clearpage

\begin{deluxetable}{ccccc}
\footnotesize
\tablecaption{Observations. \label{tbl-1}}
\tablewidth{0pt}
\tablehead{
\colhead{ObsId\tablenotemark{a}} & 
\colhead{Start time\tablenotemark{b}} &
\colhead{Stop time\tablenotemark{b}} & 
\colhead{Exposure\tablenotemark{c}} &
\colhead{Count rate\tablenotemark{d}} 
}
\startdata
20330-02-01-00 & 1996-12-29 09:36:10 & 1996-12-29 11:34:13 & 4656 & $2.56\pm0.05$ \\
20330-02-02-00 & 1997-02-07 04:28:07 & 1997-02-07 07:08:14 & 3744 & $2.23\pm0.06$ \\
20330-02-03-00 & 1997-03-04 05:42:23 & 1997-03-04 07:48:13 & 4112 & $2.01\pm0.06$ \\
20330-02-04-00 & 1997-03-21 22:24:36 & 1997-03-22 01:14:13 & 4064 & $2.03\pm0.06$ \\
20330-02-05-00 & 1997-05-08 21:45:36 & 1997-05-09 00:05:13 & 2640 & $4.00\pm0.08$ \\
20330-02-06-00 & 1997-05-30 00:20:35 & 1997-05-30 02:20:13 & 3216 & $4.77\pm0.07$ \\
20330-02-07-00 & 1997-06-27 20:30:56 & 1997-06-27 23:25:13 & 2464 & $6.50\pm0.08$ \\
20330-02-08-00 & 1997-06-28 20:17:10 & 1997-06-28 23:29:13 & 3616 & $7.15\pm0.07$ \\
20330-02-09-00 & 1997-06-29 23:19:33 & 1997-06-30 01:36:13 & 3792 & $9.06\pm0.07$ \\
20330-02-10-00 & 1997-07-03 21:40:20 & 1997-07-04 00:01:14 & 4384 & $8.28\pm0.07$ \\
20330-02-11-00 & 1997-07-04 20:01:16 & 1997-07-04 22:04:14 & 3568 & $10.60\pm0.08$ \\
20330-02-12-00 & 1997-07-12 00:06:59 & 1997-07-12 03:21:14 & 3760 & $10.61\pm0.08$ \\
\enddata

%% Text for table notes should follow after the \enddata but before
%% the \end{deluxetable}. Make sure there is at least one \tablenotemark
%% in the table for each \tablenotetext.

\tablenotetext{a}{Observational Identification}
\tablenotetext{b}{Coordinated Universal Time}
\tablenotetext{c}{Approximate exposure time (seconds)}
\tablenotetext{d}{Count s$^{-1}$, 2-10 keV}

\end{deluxetable}

\clearpage

\begin{deluxetable}{cccccccccc} 
\tablecolumns{10} 
\tablewidth{0pc} 
\tablecaption{Absorbed power-law plus line fits. \label{tbl-2}} 
\tablehead{ 
\colhead{} & \multicolumn{4}{c}{Model 2} & \colhead{} & 
\multicolumn{4}{c}{Model 3} \\ 
\cline{2-5} \cline{7-10} \\ 
\colhead{Dataset} & \colhead{$A$\tablenotemark{a}} &
\colhead{$\Gamma$\tablenotemark{b}} & \colhead{$N_{\rm
H}$\tablenotemark{c}} & \colhead{$F_{\rm X}$\tablenotemark{d}} &
\colhead{} & \colhead{$A$\tablenotemark{a}} &
\colhead{$\Gamma$\tablenotemark{b}} & \colhead{$N_{\rm
H}$\tablenotemark{c}} & \colhead{$F_{\rm X}$\tablenotemark{d}}}
\startdata 
1 & 4.0 & $1.58^{+0.03-0.10}_{-0.03+0.09}$ & $22.7^{+4.1-1.3}_{-3.4-1.5}$ & 0.8 &
& 1.3 & $1.15^{+0.08-0.08}_{-0.09+0.07}$ & $9.8^{+0.5-1.4}_{-0.5+1.2}$ & 0.8 \\
2 & 4.3 & \nodata & $31.8^{+5.5-2.8}_{-4.6-1.5}$ & 0.7 &
& 0.7 & $0.92^{+0.09-0.12}_{-0.10+0.12}$ & \nodata & 0.7 \\
3 & 6.2 & \nodata & $15.7^{+2.0-1.8}_{-1.5-0.5}$ & 1.5 & 
& 3.5 & $1.35^{+0.06-0.12}_{-0.06+0.12}$ & \nodata & 1.5 \\
4 & 9.3 & \nodata & $10.4^{+0.9-1.2}_{-0.6-0.6}$ & 2.6 &
& 9.1 & $1.57^{+0.04-0.08}_{-0.04+0.07}$ & \nodata & 2.5 \\
5 & 9.9 & \nodata & $11.3^{+1.0-1.4}_{-0.6-0.8}$ & 2.7 &
& 8.6 & $1.52^{+0.04-0.11}_{-0.04+0.11}$ & \nodata & 2.6 \\
6 & 11.8 & \nodata & $8.6^{+0.8-1.2}_{-0.6-0.7}$ & 3.5 &
& 15.1 & $1.68^{+0.05-0.08}_{-0.04+0.08}$ & \nodata & 3.5 \\
\enddata 

%% Text for table notes should follow after the \enddata but before
%% the \end{deluxetable}. Make sure there is at least one \tablenotemark
%% in the table for each \tablenotetext.

\tablenotetext{a}{Flux of the unabsorbed power-law continuum at 1 keV
in units of $10^{-3}$ photons cm$^{-2}$ sec$^{-1}$}
\tablenotetext{b}{Photon spectral index}
\tablenotetext{c}{Measured column expressed in units of $10^{22}$ cm$^{-2}$}
\tablenotetext{d}{Flux expressed in units of $10^{-11}$ erg cm$^{-2}$ 
s$^{-1}$}

\tablecomments{Two kinds of errors are given here, and in subsequent
tables.  The first one is the statistical error (see text).  The
second one is an estimate of the uncertainty due to the systematic
error in the background subtraction.  The results from decreasing and
increasing the background are given by the superscript and subscript
values respectively.}

\end{deluxetable} 

\clearpage

\begin{deluxetable}{cccccccccc} 
\tablecolumns{10} 
\tablewidth{0pc} 
\tablecaption{Spectral fits with the Compton reflection spectrum. \label{tbl-3}} 
\tablehead{ 
\colhead{} & \multicolumn{4}{c}{Model 4} & \colhead{} & 
\multicolumn{4}{c}{Model 5} \\ 
\cline{2-5} \cline{7-10} \\ 
\colhead{Dataset} & \colhead{$A$\tablenotemark{a}} &
\colhead{$\Gamma$\tablenotemark{b}} & \colhead{$N_{\rm
H}$\tablenotemark{c}} & \colhead{$F_{\rm X}$\tablenotemark{d}} &
\colhead{} & \colhead{$A$\tablenotemark{a}} &
\colhead{$\Gamma$\tablenotemark{b}} & \colhead{$N_{\rm
H}$\tablenotemark{c}} & \colhead{$F_{\rm X}$\tablenotemark{d}}}
\startdata 
1 & 4.2 & $1.69^{+0.07-0.03}_{-0.07+0.06}$ & $23.1^{+4.0-2.3}_{-3.1-0.4}$ & 0.8 &
& 4.1 & $1.65^{+0.05-0.01}_{-0.05+0.05}$ & $23.0^{+4.9-2.2}_{-4.2-2.9}$ & 0.7 \\
2 & 4.4 & \nodata & $31.9^{+5.4-2.0}_{-4.3-1.8}$ & 0.6 &
& 4.3 & \nodata & $32.4^{+6.8-2.3}_{-6.2-3.6}$ & 0.6 \\
3 & 7.0 & \nodata & $16.3^{+2.0-0.4}_{-1.6-0.8}$ & 1.5 &
& 6.7 & \nodata & $16.8^{+2.2-1.1}_{-2.1-2.4}$ & 1.4 \\
4 & 10.1 & \nodata & $11.0^{+1.0-0.2}_{-0.6-0.4}$ & 2.7 &
& 10.3 & \nodata & $11.6^{+1.0-1.0}_{-1.0-1.3}$ & 2.6 \\
5 & 11.6 & \nodata & $11.9^{+1.0-0.1}_{-0.7-0.5}$ & 2.8 &
& 11.1 & \nodata & $12.7^{+1.0-1.0}_{-1.1-1.4}$ & 2.6 \\
6 & 14.0 & \nodata & $9.1^{+0.6-0.1}_{-0.6-0.4}$ & 3.8 &
& 13.3 & \nodata & $9.8^{+0.9-0.8}_{-0.7-0.9}$ & 3.6 \\
\enddata 

%% Text for table notes should follow after the \enddata but before
%% the \end{deluxetable}. Make sure there is at least one \tablenotemark
%% in the table for each \tablenotetext.

\tablenotetext{a}{Flux of the unabsorbed power-law continuum at 1 keV
in units of $10^{-3}$ photons cm$^{-2}$ sec$^{-1}$}
\tablenotetext{b}{Photon spectral index}
\tablenotetext{c}{Measured column expressed in units of $10^{22}$ cm$^{-2}$}
\tablenotetext{d}{Flux expressed in units of $10^{-11}$ erg cm$^{-2}$
s$^{-1}$}

\end{deluxetable} 

\clearpage

\begin{deluxetable}{cccccccc} 
\tablecolumns{5} 
\tablewidth{0pc} 
\tablecaption{Partial coverer fit. \label{tbl-4}} 
\tablehead{ 
\colhead{} & \multicolumn{5}{c}{Model 6} & \\ 
\cline{2-7} \\ 
\colhead{Dataset} & \colhead{$A$\tablenotemark{a}} &
\colhead{$\Gamma$\tablenotemark{b}} & \colhead{$N_{\rm
H1}$\tablenotemark{c}} & \colhead{$C_{\rm F}$\tablenotemark{d}}
& \colhead{$N_{\rm H2}$\tablenotemark{c}} 
& \colhead{$F_{\rm X}$\tablenotemark{e}}}

\startdata 
1 & 7.7 & $1.77^{+0.10-0.07}_{-0.05+0.07}$ & $111^{+52+113}_{-32-54}$ & $0.24^{+0.03+0.11}_{-0.07-0.01}$ & $23.4^{+4.0+0.0}_{-3.6-2.0}$ & 0.8 \\
2 & 8.0 & \nodata & \nodata & \nodata & $31.9^{+5.4-1.1}_{-5.1-1.6}$ & 0.6 \\
3 & 11.8 & \nodata & \nodata & \nodata & $16.8^{+2.0-0.7}_{-1.9-0.3}$ & 1.5 \\
4 & 17.5 & \nodata & \nodata & \nodata & $11.5^{+0.9-0.4}_{-0.7-0.2}$ & 2.7 \\
5 & 18.8 & \nodata & \nodata & \nodata & $12.5^{+1.1-0.5}_{-0.7-0.1}$ & 2.8 \\ 
6 & 22.3 & \nodata & \nodata & \nodata & $9.7^{+0.8-0.5}_{-0.6-0.1}$ & 3.8 \\
\enddata 

%% Text for table notes should follow after the \enddata but before
%% the \end{deluxetable}. Make sure there is at least one \tablenotemark
%% in the table for each \tablenotetext.

\tablenotetext{a}{Flux of the unabsorbed power-law continuum at 1 keV
in units of $10^{-3}$ photons cm$^{-2}$ sec$^{-1}$}
\tablenotetext{b}{Photon spectral index}
\tablenotetext{c}{Measured column expressed in units of $10^{22}$ cm$^{-2}$}
\tablenotetext{d}{Covered fraction}
\tablenotetext{e}{Flux expressed in units of $10^{-11}$ erg cm$^{-2}$
s$^{-1}$}

\end{deluxetable} 


\begin{thebibliography}{}
\bibitem[Anders \& Grevesse(1989)]{ag89} Anders, E., \& Grevesse,
    N. 1989, Geochimica et Cosmochimica Acta, 53, 197
\bibitem[Antonucci \& Miller(1985)]{am85} Antonucci, R. R. J., \&
    Miller, J. S. 1985, \apj, 297, 621
\bibitem[Antonucci(1993)]{ant93} Antonucci, R. 1993, \araa, 31, 473
\bibitem[Arnaud(1996)]{arn96} Arnaud, K. A. 1996, in ASP Conf. Proc. 101, 
    Astronomical Data Analysis Software and Systems V, ed. G. Jacoby \& 
    J. Barnes (San Francisco: ASP), 17
\bibitem[Awaki et al.(1990)]{awa90} Awaki, H., Koyama, K., Kunieda,
    H., \& Tawara, Y. 1990, \nat, 346, 544
\bibitem[Awaki et al.(1991)]{awa91} Awaki, H., Koyama, K., Inoue, H.,
    \& Halpern, J. P. 1991, \pasj, 43, 195
\bibitem[Ba{\l}uci{\'n}ska-Church \& McCammon(1992)]{bm92}
    Ba{\l}uci{\'n}ska-Church, M., \& McCammon, D. 1992, \apj, 400, 699
\bibitem[Bevington \& Robinson(1992)]{br92} Bevington, P. R., \&
    Robinson, D. K. 1992, Data Reduction and Error Analysis for the 
    Physical Sciences, McGraw-Hill, New York
\bibitem[Butcher et al.(1997)]{but97} Butcher, J. A., Stewart, G. C.,
    Warwick, R. S., Fabian, A. C., Carrera, F. J., Barcons, X., Hayashida,
    K., Inoue, H., Kii, T. 1997, \mnras, 291, 437
\bibitem[Cappi et al.(1999)]{cap99} Cappi, M., Bassani, L., Comastri,
    A., et al. 1999, \aap, 344, 857
\bibitem[de Vaucouleurs et al.(1991)]{dev91} de Vaucouleurs, G., de
    Vaucouleurs, A., Corwin, H., Buta, R. J., Paturel, G., \& Fouqu,
    P. 1991, Third Reference Catalogue of Bright Galaxies (Berlin:
    Springer)
\bibitem[Done et al.(1996)]{don96} Done, C., Madejski, G. M., \&
    Smith, D. A. 1996, \apj, 463, L63
\bibitem[Georgantopoulos et al.(1999)]{geo99} Georgantopoulos, I.,
    Papadakis, I., Warwick, R. S., Smith, D. A., Stewart, G. C., \&
    Griffiths, R. G. 1999, \mnras, 307, 815
\bibitem[George \& Fabian(1991)]{gf91} George, I. M., \& Fabian,
    A. C. 1991, \mnras, 249, 352
\bibitem[Ghisellini et al.(1994)]{ghi94} Ghisellini, G, Haardt, F.,
    Matt, G. 1994, \mnras, 267, 743
\bibitem[Glasser, Odell, \& Seufert(1994)]{gla94} Glasser, C. A.,
    Odell, C. E., \& Seufert, S. E. 1994, IEEE Trans. Nucl. Sci., 41, 1343
\bibitem[Gondek et al.(1996)]{gon96} Gondek, D., Zdziarski, A. A.,
    Johnson, W. N., George, I. M., McNaron-Brown, K., Magdziarz, P.,
    Smith, D., \& Gruber, D. E. 1996, \mnras, 282, 646
\bibitem[Gruber et al.(1996)]{gru96} Gruber, D. E., Blanco, P. R.,
    Heindl, W. A., Pelling, M. R., Rothschild, R. E., \& Hink, P. L. 1996,
    A\&ASS, 120, 641
\bibitem[Guainazzi et al.(2000)]{gua00} Guainazzi, M., Matt, G.,
    Brandt, W. N., Antonelli, L. A., Barr, P., \& Bassani, L. 2000, \aap,
    356, 463
\bibitem[Haardt \& Maraschi(1991)]{hm91} Haardt, F., \& Maraschi,
    L. 1991, \apjl, 380, L51
\bibitem[Haardt \& Maraschi(1993)]{hm93} Haardt, F., \& Maraschi,
    L. 1993, \apj, 413, 680
\bibitem[Haardt et al.(1994)]{haa94} Haardt, F., Maraschi, L., \&
    Ghisellini, G. 1994, \apj, 432, L95
\bibitem[Ho et al.(1997a)]{ho97a} Ho, L. C., Filippenko, A. V.,
    Sargent, W. L. W., \& Peng, C. Y. 1997a, \apjs, 112, 391
\bibitem[Ho et al.(1997b)]{ho97b} Ho, L. C., Filippenko, A. V., \&
    Sargent, W. L. W. 1997b, \apjs, 112, 315
\bibitem[Iwasawa et al.(1993)]{iwa93} Iwasawa, K., Koyama, K., Awaki,
    H., Kunieda, H., Makishima, K., Tsuru, T., Ohashi, T., \& Nakai,
    N. 1993, \apj, 409, 155
\bibitem[Iwasawa et al.(1994)]{iwa94} Iwasawa, K., Yaqoob, T., Awaki,
    H., \& Ogasaka, Y. 1994, \pasj, 46, L167
\bibitem[Jahoda et al.(1996)]{jah96} Jahoda, K., Swank, J. H., Giles,
    A. B., Stark, M. J., Strohmayer, T., Zhang, W., \& Morgan, E. 1996,
    SPIE, 2808, 59
\bibitem[Kallman \& McCray(1982)]{km82} Kallman, T. R., \& McCray,
    R. 1982, \apjs, 50, 263
\bibitem[Krolik \& Vrtilek(1984)]{kv84} Krolik, J. H., \& Vrtilek,
    J. M. 1984, \apj, 279, 521
\bibitem[Krolik \& Begelman(1986)]{kb86} Krolik, J. H., \& Begelman,
    M. C. 1986, \apjl, 308, L55
\bibitem[Krolik et al.(1994)]{kro94} Krolik, J. H., Madau, P., \&
    \.{Z}ycki, P. T. 1994, \apj, 420, L57
\bibitem[Leahy \& Creighton(1993)]{lc93} Leahy, D. A., \& Creighton,
    J. 1993, \mnras, 263, 314
\bibitem[Levine et al.(1996)]{lev96} Levine, A. M., Bradt, H., Cui,
    W., Jernigan, J. G., Morgan, E. H., Remillard, R., Shirey, R. E., \&
    Smith, D. A. 1996, \apj, 469, L33
\bibitem[Leighly et al.(1999)]{lei99} Leighly, K. M., Halpern, J. P.,
    Awaki, H., Cappi, M., Ueno, S., \& Siebert, J. 1999, \apj, 522, 209
\bibitem[Madejski et al.(2000)]{mad00} Madejski, G. M., \.{Z}ycki,
    P. T., Done, C., Valinia, A., Blanco, P., Rothschild, R., Turek,
    B. 2000, \apjl, 535, L87
\bibitem[Magdziarz \& Zdziarski(1995)]{mz95} Magdziarz, P. \&
    Zdziarski, A. A. 1995, \mnras, 273, 837
\bibitem[Marshall et al.(1992)]{mar92} Marshall, F. E., Boldt, E. A.,
    Holt, S. S., et al. 1992, in Frontiers of X-ray Astronomy, ed. Tanaka,
    Y. \& Koyama, K. (Tokyo: Universal Academy Press), 233
\bibitem[Matt et al.(1991)]{mat91} Matt, G., Perola, G. C., \& Piro,
    L. 1991, \aap, 247, 25
\bibitem[Miller \& Goodrich(1990)]{mg90} Miller, J. S. \& Goodrich, R. W. 
    1990, \apj, 355, 456
\bibitem[Mulchaey et al.(1993)]{mul93} Mulchaey, J. S., Colbert, E.,
    Wilson, A. S., Mushotzky, R. F., \& Weaver, K. A. 1993, \apj, 414, 144
\bibitem[Mulchaey et al.(1996)]{mul96} Mulchaey, J. S., Wilson, A. S.,
    \& Tsvetanov, Z. 1996, \apjs, 102, 309
\bibitem[Nandra \& Pounds(1994)]{np94} Nandra, K., \& Pounds,
    K. A. 1994, \mnras, 268, 405
\bibitem[Netzer et al.(1998)]{net98} Netzer, H., Turner, T. J., \&
    George, I. M. 1998, \apj, 504, 680
\bibitem[Reynolds et al.(1999)]{rey99} Reynolds, C. S., Heinz, S.,
    Fabian, A. C., \& Begelman, M. C. 1999, \apj, 521, 99
\bibitem[Ryde et al.(1997)]{ryd97} Ryde, F., Poutanen, J., Svensson,
    R., Larsson, S., \& Ueno, S. 1997, \aap, 328, 69
\bibitem[Simpson et al.(1996)]{sim96} Simpson, C., Mulchaey, J. S.,
    Wilson, A. S., Ward, M. J., \& Alonso-Herrero, A. 1996, \apj, 457, L19
\bibitem[Smith \& Done(1996)]{sd96} Smith, D. A. \& Done, C. 1996,
    \mnras, 280, 355
\bibitem[Svensson(1996)]{sve96} Svensson, R. 1996, \aaps, 120, 475
\bibitem[Tanaka et al.(1995)]{tan95} Tanaka, Y., Nandra, K., Fabian,
    A. C., et al. 1995, \nat, 375, 659
\bibitem[Terashima et al.(2000)]{ter00} Terashima, Y., Ho, L. C., \&
    Ptak, A. F. 2000, \apj, 539, 161
\bibitem[Tran et al.(1992)]{tra92} Tran, H. D., Miller, J. S., \& Kay,
    L. E. 1992, \apj, 397, 452
\bibitem[Turner et al.(1997a)]{tur97a} Turner, T. J., George, I. M.,
    Nandra, K., \& Mushotzky, R. F. 1997a, \apjs, 113, 23
\bibitem[Turner et al.(1997b)]{tur97b} Turner, T. J., George, I. M.,
    Nandra, K., \& Mushotzky, R. F. 1997b, \apj, 488, 164
\bibitem[Turner et al.(2000)]{tur00} Turner, T. J., Perola, G. C.,
    Fiore, F., Matt, G., George, I. M., Piro, L., Bassani, L. 2000, \apj,
    531, 245
\bibitem[Vignali et al.(1998)]{vig98} Vignali, C., Comastri, A.,
    Stirpe, G. M., Cappi, M., Palumbo, G. G. C., Matsuoka, M., Malaguti,
    G., \& Bassani, L. 1998, \aap, 333, 411
\bibitem[Warwick et al.(1989)]{war89} Warwick, R. S., Koyama, K.,
    Inoue, H., Takano, S., Awaki, H., \& Hoshi, R. 1989, \pasj, 41, 739
\bibitem[Warwick et al.(1993)]{war93} Warwick, R. S., Sembay, S.,
    Yaqoob, T., Makishima, K., Ohashi, T., Tashiro, M., \& Kohmura,
    Y. 1993, \mnras, 265, 412
\bibitem[Xue et al.(1998)]{xue98} Xue, S.-J., Otani, C., Mihara, T.,
    Cappi, M., \& Matsuoka, M. 1998, \pasj, 50, 519
\bibitem[Yaqoob et al.(1993)]{yaq93} Yaqoob, T., Warwick, R. S.,
    Makino, F., Otani, C., Sokoloski, J. L., Bond, I. A., \& Yamauchi,
    M. 1993, \mnras, 262, 435
\bibitem[Zdziarski et al.(1995)]{zdz95} Zdziarski, A. A., Johnson,
    W. N., Done, C., Smith, D., \& McNaron-Brown, K. 1995, \apj, 438, L63
\bibitem[Zhang et al.(1993)]{zha93} Zhang, W., Giles, A. B., Jahoda,
    K., Soong, Y., Swank, J. H., \& Morgan, E. 1993, SPIE, 2006, 324
\end{thebibliography}
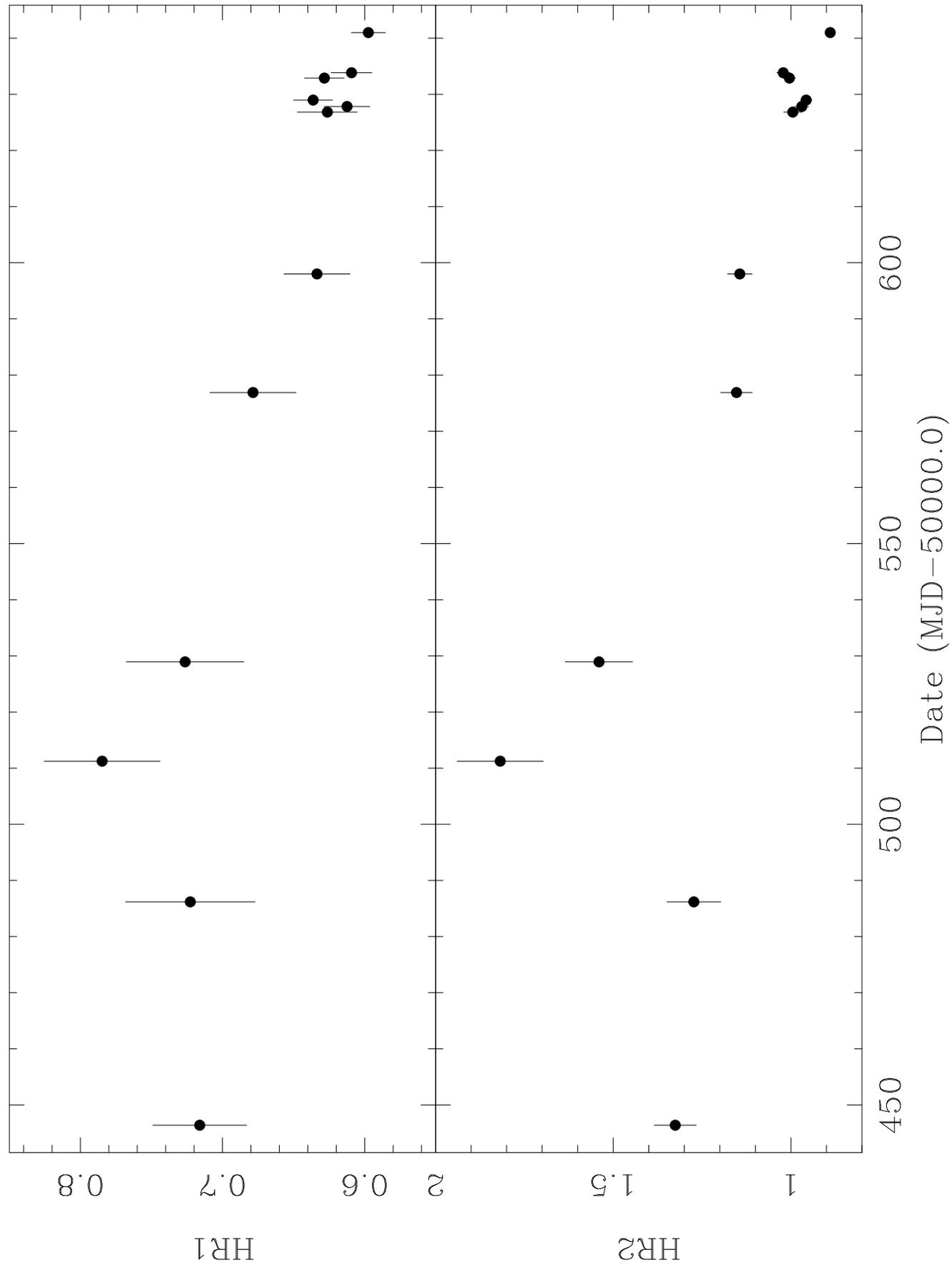
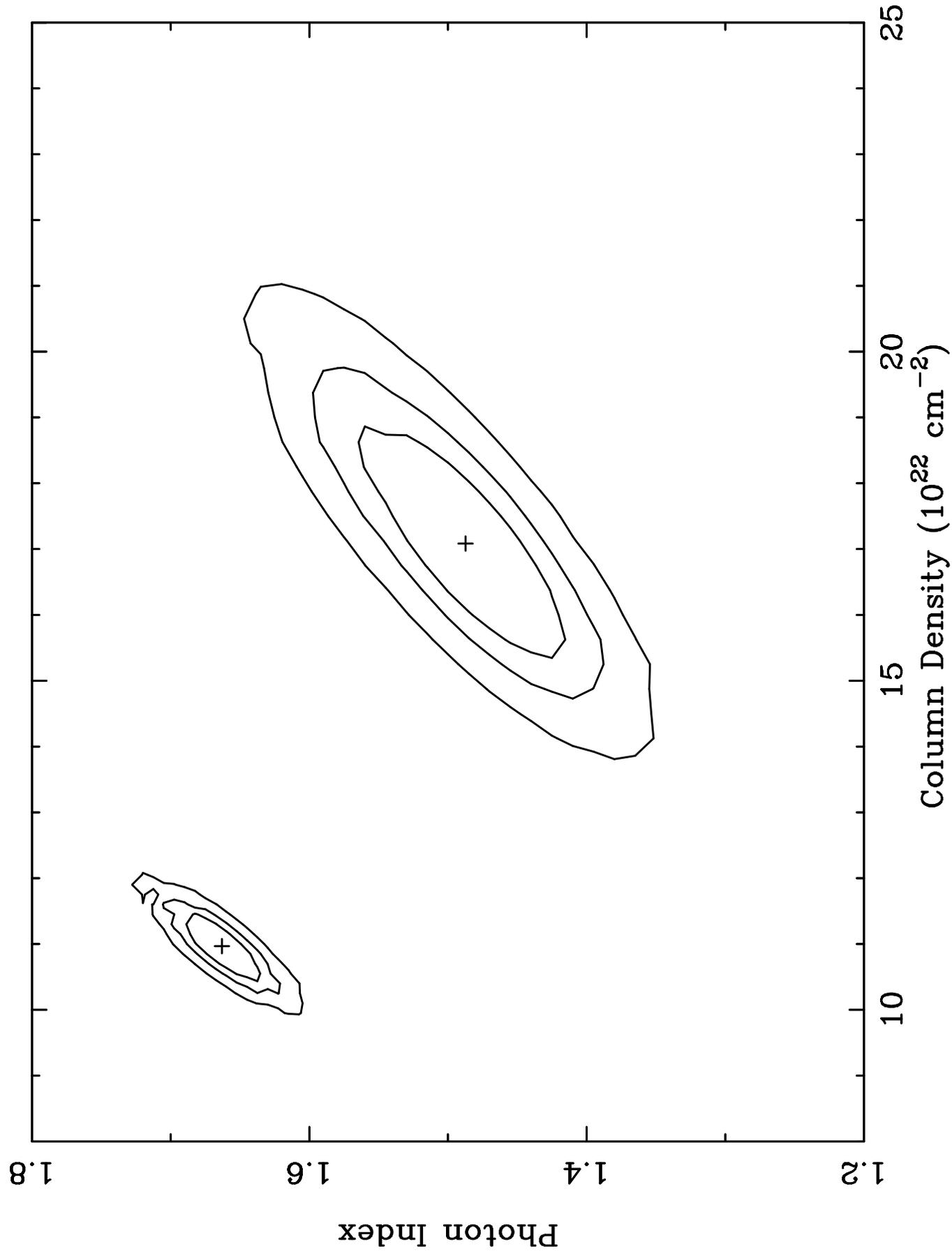
\end{document}